# THE COSMIC DUST ANALYSER ONBOARD CASSINI:
# TEN YEARS OF DISCOVERIES

R. Srama (1,2), S. Kempf (3,13), G. Moragas-Klostermeyer (1), N. Altobelli (4), S. Auer (5), U. Beckmann (6), S. Bugiel (2,1), M. Burton (7), T. Economomou (8), H. Fechtig (1), K. Fiege (1,3), S. F. Green (9), M. Grande (10), O. Havnes (11), J. K. Hillier (9,19), S. Helfert (12), M. Horanyi (13), S. Hsu (1,13), E. Igenbergs (14), E. K. Jessberger (15), T. V. Johnson (7), E. Khalisi (1), H. Krüger (6), G. Matt (1), A. Mocker (1,2), P. Lamy (16), G. Linkert (1), F. Lura (17), D. Möhlmann (17), G. E. Morfill (18), K. Otto (13), F. Postberg (1,2,19), M. Roy (7), J. Schmidt (20), G. H. Schwehm (5), F. Spahn (20), V. Sterken (1,3), J. Svestka (21), V. Tschernjawski (17), E. Grün (1,13), H.-P. Röser (2)

## Abstract

The interplanetary space probe Cassini/Huygens reached Saturn in July 2004 after seven years of cruise phase. The German Cosmic Dust Analyser (CDA) was developed under the leadership of the Max Planck Institute for Nuclear Physics in Heidelberg under the support of the DLR e.V.. This instrument measures the interplanetary, interstellar and planetary dust in our solar system since 1999 and provided unique discoveries. In 1999, CDA detected interstellar dust in the inner solar system followed by the detection of electrical charges of interplanetary dust grains during the cruise phase between Earth and Jupiter. The instrument determined the composition of interplanetary dust and the nanometre sized dust streams originating from Jupiter's moon Io. During the approach to Saturn in 2004, similar streams of submicron grains with speeds in the order of 100 km/s were detected from Saturn's inner and outer ring system and are released to the interplanetary magnetic field. Since 2004 CDA measured more than one million dust impacts characterizing the dust environment of Saturn. The instrument is one of three experiments which discovered the active ice geysers located at the south pole of Saturn's moon Enceladus in 2005. Later, a detailed compositional analysis of the water ice grains in Saturn's E ring system lead to the discovery of large reservoirs of liquid water (oceans) below the icy crust of Enceladus. Finally, the determination of the dust-magnetosphere interaction and the discovery of the extended E ring (at least twice as large as predicted) allowed the definition of a dynamical dust model of Saturn's E ring describing the observed properties. This paper summarizes the discoveries of a ten year story of success based on reliable measurements with the most advanced dust detector flown in space until today. This paper focuses on cruise results and findings achieved at Saturn with a focus on flux and density measurements. CDA discoveries related to the detailed dust stream dynamics, E ring dynamics, its vertical profile and E ring compositional analysis are published elsewhere (Hsu et al. 2010, Kempf et al., 2008, 2010, Postberg et al, 2008, 2009).

(1) Max Planck Institut für Kernphysik, Heidelberg
(2) IRS, Universität Stuttgart, D
(3) Universität Braunschweig, D
(4) ESAC, Madrid, ESP
(5) A&M Associates, Basye, USA
(6) MPS, Kathlenburg-Lindau, D
(7) JPL, Pasadena, USA
(8) Universität Chicago, USA
(9) Open University, Milton Keynes, U.K.
(10) University of Wales Aberystwyth, U.K.
(11) Universität Tromso, NOR
(12) Helfert Informatik GmbH & Co KG, Mannheim, D
(13) LASP/Universität Colorado, Boulder, USA
(14) Universität München, D
(15) Universität Münster, D
(16) LAS, Marseille, F
(17) DLR Berlin, D
(18) MPE Garching, D
(19) Universität Heidelberg,
(20) Universität Potsdam, D
(21) Prag Observatorium, CR (23)

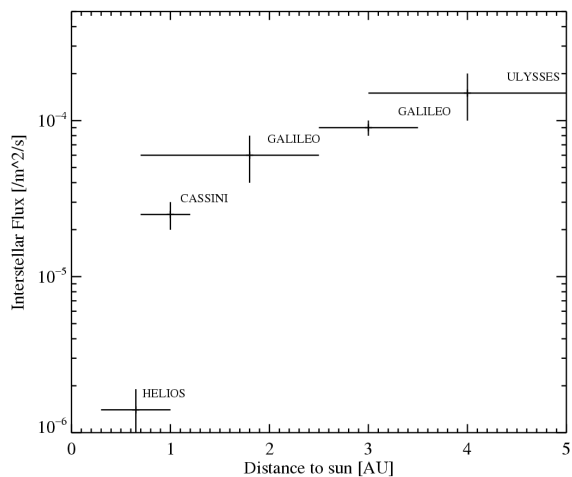

Figure 1: The ISD flux measured by various interplanetary spacecraft. The radiation pressure of the Sun and interaction of the charged grains with the interplanetary magnetic field prevent small particles from penetrating the inner heliosphere.

## Introduction

The NASA-ESA space mission Cassini-Huygens reached the ring planet Saturn in July 2004 after seven years of interplanetary cruise and is investigating Saturn with its magnetosphere, its moons and ring system. The Cosmic Dust Analyser (CDA), a joint development of the MPI Nuclear Physics in Heidelberg and the DLR Berlin, provides unique insights into the dusty environment of the large planet. The CDA determines the speed (1 - 100 km $s^{-1}$), mass ($10^{-15}$ - $10^{-9}$ g), electric charge (1 fC - 1 pC) and elemental composition (m/$\Delta$m = 20 - 50) of individual micrometeoroids and derives dust densities between $10^{-9}$ $m^{-3}$ and 10 $m^{-3}$. The instrument has a mass of 17 kg, consumes 12 W and its nominal data rates are between 128 and 4192 bps (Srama 2000, 2004). Many years ago, its time-of-flight mass spectrometer was manufactured under strong contamination guidelines and procedures enabling the science team today to derive invaluable compositional information of Saturn's ring particles which cannot be achieved by other means and instruments (Postberg et al., 2009). A specially designed articulation mechanism allows the instrument to rotate about one axis and to independently track the dust RAM direction onboard the three axis stabilised Cassini spacecraft. A international team of scientists and engineers under the leadership of the MPIK in Heidelberg and, since 2011, at the University of Stuttgart, performs science planning, operations including commanding and instrument monitoring, data processing and science analysis.

The instrument consists of an Impact Ionisation Detector (IID, Srama et al., 2004), a Chemical Analyser (CA) and a High Rate Detector (HRD). Various subsystems are based upon charge induction, impact ionisation, time-of-flight mass spectrometry and depolarisation of PVDF foils (Simpson et al., 1985). This combination of subsystems provides reliable information about individual dust impacts in Saturn's magnetosphere with its variable plasma and dust densities.

## Cruise Science

### Interstellar dust in the inner solar system

Almost 20 years ago, interstellar dust was identified inside our planetary system by the dust detector onboard the Ulysses spacecraft [Grün et al., 1993]. A flow of µm-sized interstellar grains has been identified at a distance of up to about five AU from the Sun. The observed flux was $1.5 \times 10^{-4}$ $m^{-2}$ $s^{-1}$ of particles with a mean mass of $3 \times 10^{-13}$ g giving a mass flux of $5 \times 10^{-17}$ g $m^{-2}$ $s^{-1}$. The results showed, that interstellar dust enters the Solar System with 26 km $s^{-1}$ and its upstream direction of 259° longitude and +8° latitude was found to be compatible with the direction of the interstellar gas [Landgraf, 2000]. This finding lead to the integration of a special interstellar dust (ISD) observation period around 1 AU solar distance in the early cruise phase of Cassini in 1999. This observation with very limited spacecraft resources provided the first unique discovery of CDA, the in-situ measurement of individual ISD grains in the inner solar system [Altobelli et al., 2005b].[1] An ISD flux of approximately 2.5x10$^{-5}$ m$^{-2}$ s$^{-1}$ was detected between 0.7 and 1.3 AU heliocentric distance of grains with masses compatible with values determined by Ulysses of 3x10$^{-16}$ kg (Fig. 1). The decrease of the ISD flux at close heliocentric distances can be explained by the radiation pressure filtering effect, preventing the smaller grains to reach the innermost regions of the Solar System. On the other hand, the spatial density of big ISD grains is enhanced by gravitational focusing at close heliocentric distances. The deep penetration of ISD into the solar system started a series of discoveries of the most advanced dust detector ever flown on an interplanetary mission.

**Electrostatic charges of interplanetary dust particles**

Cosmic dust particles are embedded in the heliosphere and interact with UV light of the Sun and solar wind particles. They are affected by a variety of charging mechanisms leading normally to an equilibrium potential at the grain surface of about +5V. UV photoelectron emission dominates the charging process in regions with low plasma densities like the interplanetary space. Other contributing charging processes are sticking and penetration of plasma particles, and secondary electron emission due to the bombardment of highly energetic plasma particles.

The CDA instrument has the capability to determine the electrical charge of incident dust grains with a sensitivity of approximately 1 fC. Between 1 AU and 2.1 AU heliocentric distance CDA registered six impacts showing a clear charge signal of interplanetary dust particles (IDP) (Fig. 2). This was the first unambiguous detection of electrostatic charges carried by dust particles in interplanetary space and a detailed description can be find in Kempf et al., 2004. The detector geometry with two inclined charge sensing grids allows an accurate determination of the particle speed, electrostatic charge, mass, and directionality. The dust charges varied between 1.3 and 5.4 fC corresponding to particles masses between 1.3x10$^{-13}$ and 9.5x10$^{-12}$ kg. Here, for the first time, a reliable grain mass determination was derived under the assumption of a grain surface potential of +5 V. The particle properties are in agreement with the interplanetary dust model predictions and the grain speeds are close to the predicted speed of circular Kepler velocities of micrometeoroids at the Cassini location during the time of detection.

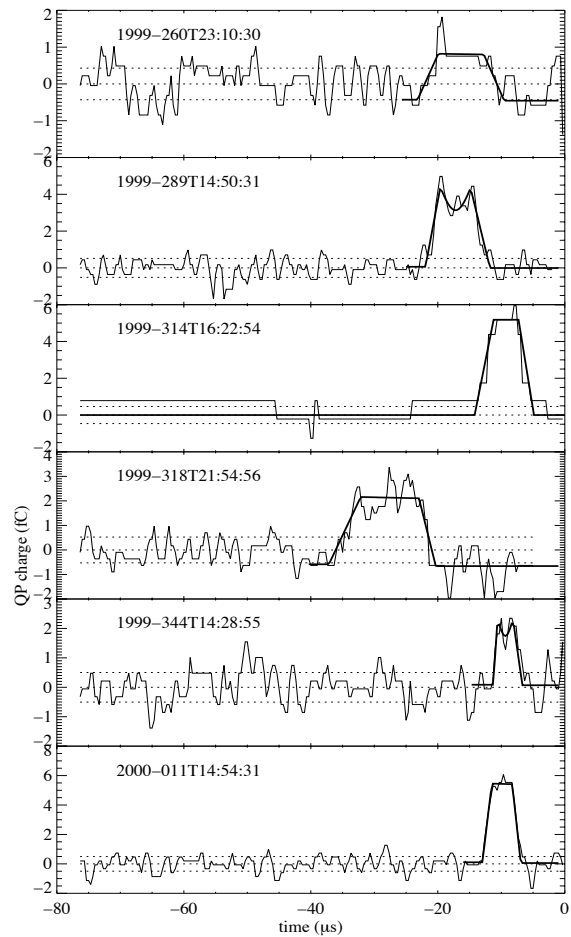

Figure 2: Measurement signals of the CDA entrance grid channel (QP) of IDPs carrying electrostatic charges between 1.3 and 5.4 fC recorded during the interplanetary cruise in 1999 and 2000 between Earth and Jupiter. The reconstructed charge features are indicated by thick lines (Kempf et al., 2004). The impact speeds are 15 km s-1, 21 km s-1, 34 km s-1, 13 km s-1, 45 km s-1 and 34 km s-1.

---

[1] The ISD observations were performed using extremely low data rates. Highly processed and compressed science data were transported by the instrument housekeeping data channel. Furthermore, no spacecraft pointing changes were allowed.

## Composition of Interplanetary Dust Particles

Early in Cassini's interplanetary cruise phase, before the flyby of the Jovian system, the CDA's Chemical Analyser subsystem detected two interplanetary dust particles (Hillier et al., 2007). During this time the instrument was not in full science mode and serendipitous science data was returned with engineering and housekeeping data. The first IDP spectrum was recorded on May 27th, 1999, when Cassini was 0.89 AU from the Sun. The second spectrum was recorded at a heliocentric distance of 1.87 AU, on the 10th November 1999.

Laboratory calibration of the CDA flight spare using the Van de Graaff dust accelerator in Heidelberg indicates that the first particle had a mass of $9[+55,-18] \times 10^{-14}$ kg and struck CDA at a speed of $18+/-10$ km s$^{-1}$. The second particle had a mass of $1.4[+1.9,-0.8] \times 10^{-12}$ kg and impacted at a speed of $7.7 \pm 4.6$ km s$^{-1}$.

The time of flight mass spectra for both impacts are shown in Fig. 3. The mass resolution of the spectra is reduced by special instrument settings, due to bandwidth limitations. Calibration of the spectrum for impact 1 was complicated by the triplet of features between 4.5 and 5.2 μs. The Rhodium peak (ions from the instrument target) is usually used for calibration onto a mass scale. Comparison of the first impact spectrum with the second allowed the Rhodium peak to be reliably identified and mass scales for both spectra to be derived. Numerical ion trajectory modelling, comparison with laboratory calibration spectra and solar system elemental abundances indicate that the spectrum of impact 1 exhibits features of fast non-thermal ($Fe_f$ and $Rh_f$) and slow thermal ($Fe_s$, $Rh_s$) ions[2]. Both grains were found to be similar in composition, predominantly Fe with surprisingly few traces of silicates. The apparent compositions of the particles, together with their detection locations and impact velocities indicate that they are in asteroidal-type orbits (Aten and Apollo) although cometary-type orbital solutions are possible.

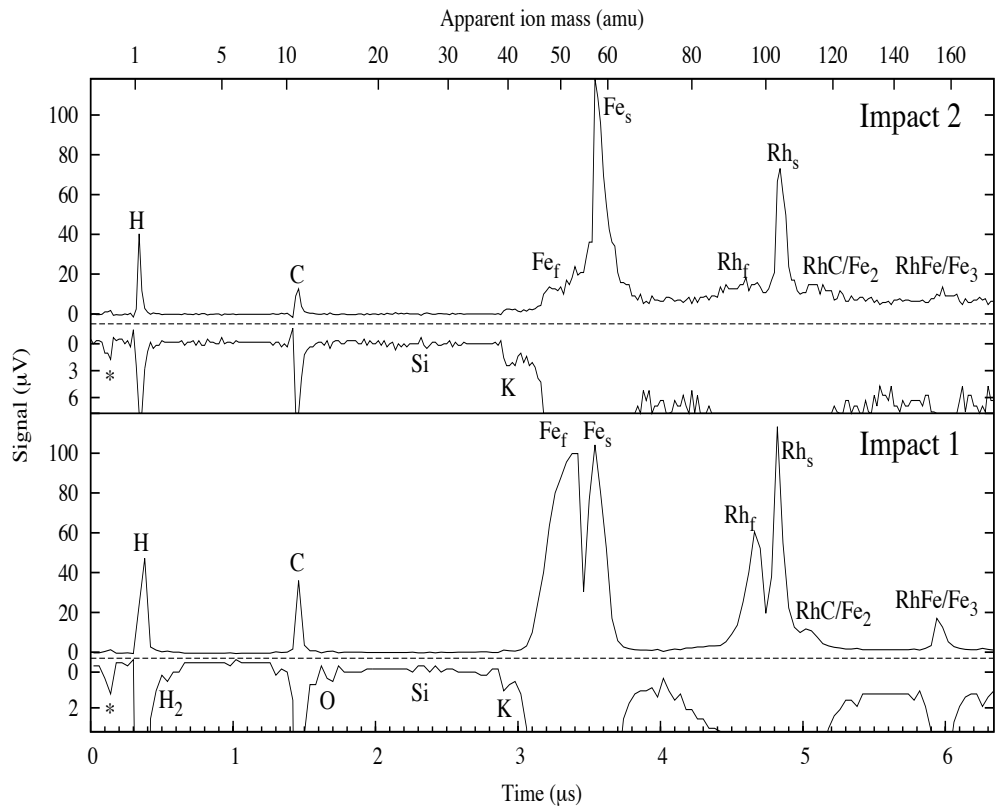

Figure 3: CDA Chemical Analyser time of flight mass spectra of two interplanetary dust particles (Hillier et al., 2007). The peaks labeled * are an instrument artefact. For clarity the spectra are shown inverted and magnified. The magnified view of the Impact 1 spectrum is magnified by a factor of 9, that of Impact 2 by a factor of 4.5. Carbon and Hydrogen are common contaminants of the Rhodium impact target but may also occur in the interplanetary dust particles.

**Composition of Jovian dust streams**

Streams of nanometre sized dust grains originating from Jupiter's moon Io were discovered and investigated by the dust instruments onboard the interplanetary space missions Ulysses and Galileo (Grün et al., 1993, Grün et al., 1996, Krüger et al., 2003, Krüger et al., 2006, Graps et al., 2000). These particles are released by Io's volcanoes, become charged in the plasma torus around Jupiter and are accelerated to speeds as high as 400 km s$^{-1}$ leaving the Jovian system before coupling to the interplanetary magnetic field in interplanetary space.

The Cassini flyby of Jupiter in 2000 provided the unique opportunity to investigate the Jovian dust stream phenomenon by a sophisticated dust detector. Between day 248 in year 2000 and day 165 in year 2001 the CDA sensor recorded 7283 stream particle impacts including 836 time-of-flight (TOF) mass spectra corresponding to a Jovian distance of 1.1 AU at the inbound trajectory and about 2 AU on the outbound trajectory. Similar to Ulysses and Galileo, Cassini CDA observed bursts of high dust impact rates with durations between one and ten days. The time interval between the two strong impact bursts at day of year (DOY) 251 and DOY 266 in the year 2000 is approximately 15 days which agrees well with dust streams modulated by the IMF with a four sector structure (Kempf, 2007, Habil.). Postberg et al. (2006) analysed 287 spectra of these two major burst periods and he found nine distinct features labeled with $F_x$ (Fig. 4) in the mass spectrum with positive ions. The dominant spectral features are H+ (F1), C2+ (F2), C+ (F3), O+ (F4), Na+ (F5), Si+ (F6), S+/Cl+ (F7), K+ (F8) and Rh+ (F9). H+ and C+ are mainly attributed to contaminations and Rh+ represents the target material. Therefore sodium, chlorine, sulfur and potassium are the main particle components. Due to a clear quantitative correlation NaCl is the parent molecule of the majority of the recorded Na+ and Cl+ ions. Silicates and sulfur or sulfurous components might be another constituent of the stream particles, which have grain sizes of approximately 24 nm. From this chemical fingerprint, recorded approximately

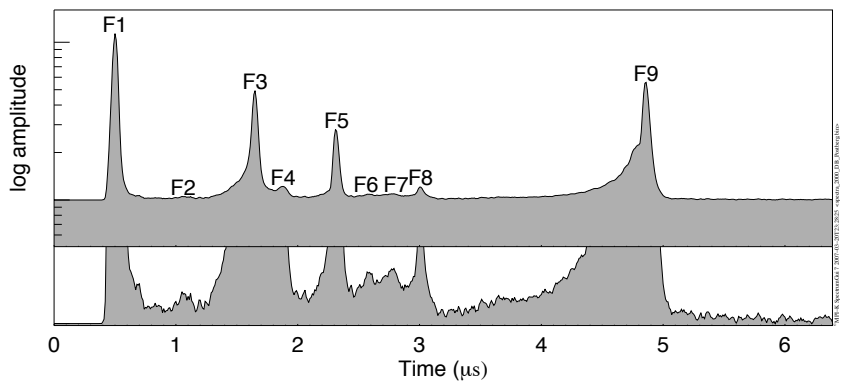

Figure 4: Dust spectrometry of Io's volcanic ash recorded more than 80 Mkm away from Jupiter. A sum spectrum of 30 TOF mass spectra of Jupiter's nanometre sized stream particles shows nine major peak features indicating a NaCl rich composition. The lower spectrum shows a stretched y-scale (Kempf, 2007).

---

[2] The peak shapes reflect the energy and angular distribution of the ion species in the impact plasma. The CDA instrument with its integrated linear TOF mass spectrometer does not compensate for different ion energies. The initial ion energy distribution is dependent on the particle impact energy and impact velocity.

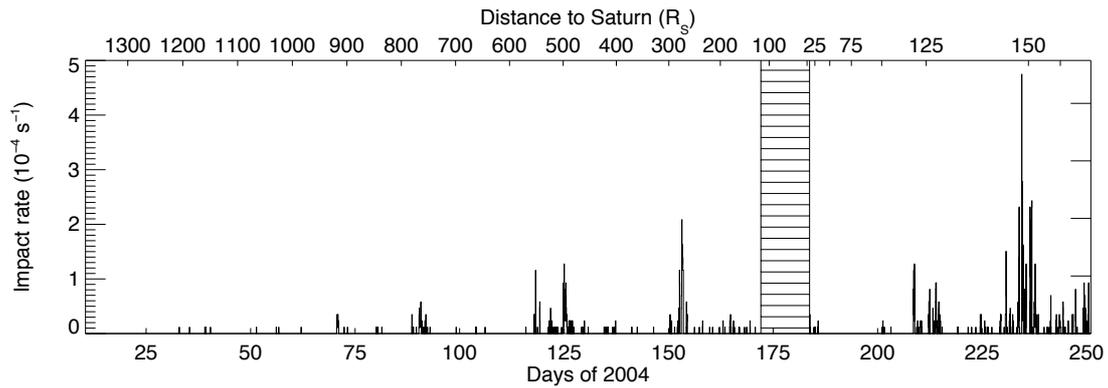

Figure 5: Discovery of stream particles by CDA originating from the Saturnian system during the approach phase to Saturn between January 10 and September 6 in 2004 (Kempf et al., 2005, Nature). CDA was powered off during the orbit insertion phase (SOI) between 20 June and 1 July indicated by the vertical stack of bars. The upper scale gives Cassini's distance to Saturn in RS. In total, 1,409 impacts were detected: 546 before SOI and 863 impacts after SOI).

100 million km away from Jupiter, Io's volcanoes are again identified as the source for the vast majority of stream particles detected. It is remarkable, that such an in-situ analysis far away from its source allow us to learn about the thermodynamic and chemical properties of the volcanoes and its plumes and even to look inside a small planetary body. Based on these observations, Postberg et al. (2006) suggested a condensation cascade inside the erupting volcanic gases around pryoclastic silicate cores. Sodium and potassium condenses prior to sulfuric compounds ($SO_2$) due to their high condensation temperatures, explaining the low contents of sulfur in the mass spectra.

## Science at Saturn

### Discovery of Saturn Stream particles

The potential mechanism to accelerate particles within Jupiter and Saturn's magnetosphere to speeds exceeding the planet's escape velocity was discussed by e.g. Horanyi et al. (2000) and by Kempf (2007). In order to escape from Saturn, positively charged grains have to be accelerated by the outward-pointing co-rotational electric field caused by Saturn's rotating magnetic field. Horanyi predicted the occurrence of tiny grains released by the Saturnian system along the ring plane. Since the magnetic field of Saturn is weaker by a factor of 20 than the magnetic field of Jupiter, the escaping grains are either slower or smaller than in the Jovian case. In order to search for such streams originating from Saturn, special observation campaigns were integrated in the approach phase to Saturn starting in January 2004. The planning efforts were rewarded immediately by the detection of faint bursts of nanometre sized stream particles starting at distances as far as 1200 $R_S$ (70 Mkm) away from Saturn (Fig. 5). The burst intensities and impact rates seemed to grow with decreasing distance to Saturn implying an origin from the Saturnian system. Before Cassini's orbit insertion 546 impacts were detected. According to the instrument calibration, the impact speed exceeded 70 km s$^{-1}$ from a direction which excludes the so-called β-meteoroids, which is in agreement with the model predictions of velocities above 100 km s$^{-1}$, and which excludes clearly bound interplanetary particles with their low relative impact speeds (<20 km s$^{-1}$). Besides the Jovian system, Saturn with its ring system is now the second source of very tiny dust particles (masses below 10$^{-21}$ kg) moving through our solar system with speeds above 100 km s$^{-1}$. Generally, it is even possible that particles from the Saturnian system enter the Jovian magnetosphere and vice versa.

What do we know about these particles besides their masses and speeds? Is it possible to derive the composition of nanometre sized micrometeoroids? The answer is yes, although by employing a special spectra processing method using an iterative method in order to magnify the weak features. The most prevalent mass lines, $C^+$, $O^+$, $Na^+$, $K^+$, $Fe^+$, $SiO_2^+$, and $Rh^+$ were identified by adding spectra of similar strength and with similar features. Then, each individual spectrum was scanned for these lines and an individual and improved mass scale was allocated. A new sum spectrum was calculated using the improved mass scales. This procedure was repeated until the resolution of the spectral lines of the sum spectrum did not improve. Finally, a consistent data set of calibrated individual mass spectra was obtained. Besides the aforementioned ions, clear evidence for $H_3O^+$, which is characteristic for water ice, was found. There is no indication, however, of water cluster ions $(H_2O)_xH_3O^+$, with x = 1,2,... typically found for impacts by water ice grains at low impact energies. About 25% of the spectra also showed N+ and a line at 18 amu ($NH_4^+$). Furthermore, there is evidence for $Si^+$ (28 amu), $SiO_2^+$ and $Fe^+$, which are the building blocks of typical minerals like olivine ($(Mg,Fe)_2SiO_4$). Carbon dominates all spectra but is probably caused by target contaminations. Four compositional types were defined according to the identified lines of each individual spectrum (Fig. 6): silicate type for spectra showing at least $Si^+$ or $SiO_2^+$, or $Fe^+$ peaks, alkali type for spectra showing at least $Na^+$ or $K^+$ peaks, water ice type for spectra showing both, $O^+$ and $H_3O^+$ peaks, and ammonium type for spectra showing both, $N^+$ and $NH_3^+$ lines. 74% of the spectra are of silicate type (32% of them simultaneously belong to the ice type), but only about 7% of the spectra are of ice type but show no silicate lines. The preponderance of spectra with silicate lines suggest that the majority of the detected grains consists of silicate material. If the water ice lines are not due to target contaminations then about a third of the silicate grains had ice mantles, whereas solid ice grains are rare.

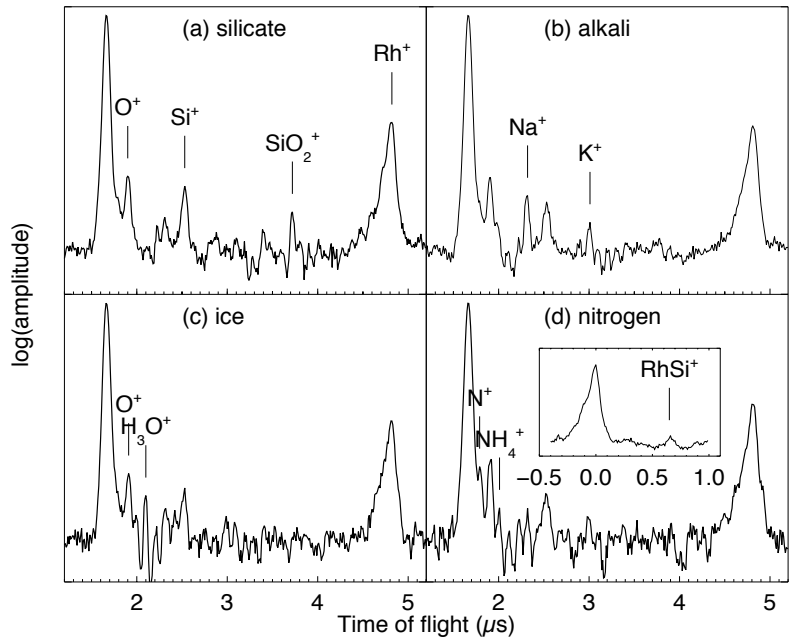

Figure 6: Co-added TOF mass spectra of the four composition types: silicates, alkali, ice and nitrogen. The CDA spectra of stream particles indicate that the majority of the nano-grains consists of silicate material, about 30% of them have mantles of water ice and/or clathrate hydrates of ammonia (Kempf et al. 2005b).

Saturn's main ring is thought to be primarily made of water ice, possibly containing clathrate hydrates of ammonia or methane and a minor amount of impurities most likely iron-bearing silicate compounds. E ring particles dominantly consist of pure water ice. Ring particles originating from Enceladus plumes also contain impurities – probably tiny silicate inclusions or organic compounds (Postberg et al., 2007). If the plume particle inclusions are silicates, then the Enceladus plumes are an important source of Saturnian stream particles. In any case, in-situ studies of stream particles originating from the inner Saturnian system allowed the first and only study of the composition of the ring system and of the moons.

**Dynamics and origin of stream particles**

The small size of stream particles generally leads to high charge-to-mass ratios such that electromagnetic forces significantly determine their dynamics. To overcome Saturn's gravity, the charge-to-mass ratio Q/m of the grain has to be sufficiently large. This constraint sets an upper limit on the radius of a grain escaping from the inner Saturnian system. There is also a lower size limit, as Q/m has to be sufficiently small that grains circle along a magnetic field line and become tied to Saturn's magnetic field. Of the first order, only positive charged particles are accelerated outwards in Saturn's magnetosphere. The grain charge results from a competition between various charging processes like collection of ions or electrons, emission of photo electrons and emission of secondary electrons. There are two regions with a low plasma density where the dust surface potential is expected to be positive (photo emission) and which are considered as stream particle origin: The outskirts of the A ring and the region outside of approximately 7 Saturn radii (Fig. 7, Kempf et al., 2005a). The dark region between 3 and 6 Saturn radii correspond to negative surface potentials. It is a dynamical barrier that prevents nanodust particles located within or inward this region to escape. Theoretically, the stochastic charging behaviour of nanodust mitigates the negative potential barrier. Nevertheless, in addition to the negative charging barrier, the sputtering of nanoparticles in this dense plasma region limits the dust lifetime and indicates that only sputtering resistant material can survive and become stream particles. This escape process has been examined by Hsu et al. (2011). Combining CDA composition analysis and the backward tracing simulation results, the authors suggested, that Saturnian stream particles are dynamically old, silicateous relics released from plasma-eroded E ring icy grains. This implies that stream particles are of the same origin as E ring dust grains - Enceladus.

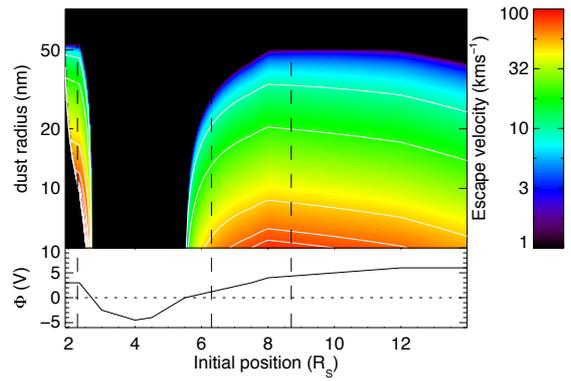

Figure 7 : Escape velocity of Saturnian stream particles as a function of the initial location (distance from Saturn) versus the dust radius shown as a colour scale (top panel). The bottom panel shows the equilibrium potential Φ used for calculating the exit speed vex. The vertical dashed lines mark the outer edge of the A ring and the location of Saturn's moons Dione and Rhea. From Kempf et al. (2005a, Nature).

Outside Saturn's magnetosphere, the dynamics of the grains are governed by their interaction with the IMF, convected by the solar wind. As the Cassini spacecraft crosses the compression regions of the Co–rotation Interaction Regions (CIRs), not only the directionality of the impacts changes with the field direction, but also the impact signal and rate vary with an increase of field strength (bursts occur). At large solar distances the azimuthal component of the IMF dominates, and thus the out-of-ecliptic component of the dust velocity should be affected most. As the inertial spacecraft velocity of about 10 km s$^{-1}$ is small compared with the dust speed, deviations of the impact direction from Saturn's location are attributed to a bending of the dust trajectories by the Lorentz force due to the IMF. Monitoring the dust streams from Saturn's magnetosphere is useful for identifying their source, and for understanding both the acceleration mechanism and the coupling between dust (including the rings) and the magnetic field and the plasma environment in the magnetosphere (Hsu et al., 2011).

## G ring encounter

The optical properties of the G ring are dominated by micron-sized dust grains, which are expected to erode quickly by sputtering. Because G ring dust is considered to be hazardous for Cassini, the spacecraft trajectory shall not cross the inner core of the G ring.

| counter | radius ($\mu$m) | impact # | counter | radius ($\mu$m) | # of counts |
|---------|-----------------|----------|---------|-----------------|-------------|
| $m_1$ | 1.2 | 356 | $M_1$ | 2.1 | 278 |
| $m_2$ | 2.4 | 55 | $M_2$ | 4.2 | 46 |
| $m_3$ | 4.9 | 3 | $M_3$ | 8.6 | 1 |
| $m_4$ | 6.7 | 0 | $M_4$ | 11.9 | 1 |

Table 1: Mean particle radius required to increment an HRD counter during the G ring encounter on 2005-248. The relative speed of the ring particles to the detector was about 8 km s−1 and the grains were assumed to be ice particles. Note that the radii given for the highest thresholds are the minimum grain size. The instrument has a small detector using a 6 μm thick foil (counter mx), and a large detector (counter Mx) using a 28 μm thick foil. Four counters are allocated to each sensor. The counters represent different signal amplitudes and trigger thresholds.

Cassini passed the outskirts of Saturn's G ring (Fig. 8). CDA's big impact ionisation target with its field-of-view (FOV) of +/-45° was sensitive to ring particles only until about 10:40 h, while the High Rate Detector (HRD, Srama et al., 2004) with its PVDF foils and its significantly larger angular FOV could detect ring particles until about 11:00 h. Thus, both detector subsystems were capable to detect ring particles. During the encounter both HRD sensor foils worked nominally. There were a significant number of impacts exceeding the higher thresholds. At 2005-248T10:38:25 the M-sensor[3] registered an event exceeding its highest threshold M4. Unfortunately, this event probably triggered a higher sensitivity of the M-sensor leading to a background noise level during the remaining tour with rates between 0.01 s-1 and 0.1 s-1. There is some evidence by laboratory experiments that the particle size was at least 100 μm because a foil damage was not observed in impact experiments with 60 μm iron particles (Kempf, 2005).

Assuming a particle moving in a gravitationally bound circular orbit, the impact speed during the crossing was approximately 8 km s−1 corresponding to grain radii exceeding 12 μm (Tab. HRD). Note, however, that this is only a lower limit for the size of the grain causing the possible damage to the M sensor. The small HRD sensor was not affected by the G ring crossing although the foil thickness of this sensor is only 6 μm.

The HRD finding of extremely large ice particles at the outer G ring rim is consistent with observations by the Cassini plasma camera MIMI, which measures the depletion of electrons in the plasma environment. Large clumps of material obscure the motion of plasma particles gyrating around the field lines of Saturn's magnetic field.

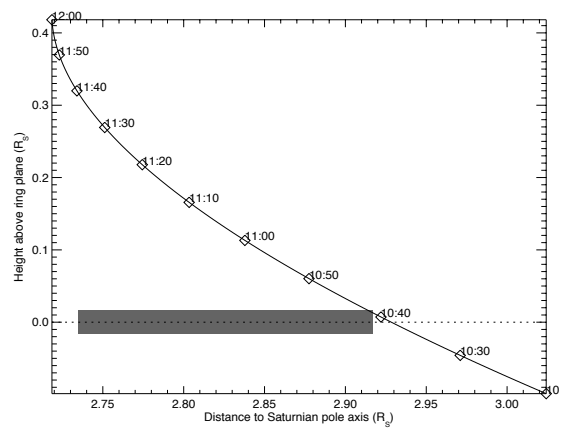

Figure 8: Spacecraft trajectory projected into a meridian plane of Saturn. The shaded area indicates the G ring assuming a vertical height of 960 km (Kempf, 2005, Techn. Report).

---

[3] 28 μm thick PVDF foil with a sensitive area of 50 cm2

## Saturn's E ring

Saturn's outer E ring is one of the most exciting astronomical objects in our solar system and it allows to study the interaction of dust, plasma, moons and the magnetosphere. The E ring is the largest planetary ring in the solar system, encompassing the icy satellites Mimas ($r_M$ = 3.07 $R_S$), Enceladus ($r_E$ = 3.95 $R_S$), Tethys ($r_T$ = 4.88 $R_S$), Dione ($r_D$ = 6.25 $R_S$), and Rhea ($r_R$ = 8.73 $R_S$). Optical remote sensing of the ring shows a bluish color of the reflected light which is typical for a narrow grain size distribution between 0.3 and 3 μm. The icy moon Enceladus was proposed early as the dominant source of ring particles since the edge-on brightness profile peaks near the moon's mean orbital distance. For the same reason, Tethys was identified as a secondary E ring particle source by de Pater et al. (2004). Photometrical models propose that for micron-sized grains the peak optical depth is τ ~ 1.6x10$^{-5}$ corresponding to about 180 grains per square centimeter (Showalter, Cuzzi, & Larson, 1991). The vertical profile of the ring was investigated by De Pater et al. (2004) by Earth-bound infrared observations in 1995. The results showed a full–width–half–maximum (FWHM) based on grain sizes dominating the optical cross section of about 9.000 km between the Mimas and the Enceladus orbit, and a minimum of 8.000 km at the distance of Enceladus. The vertical thickness of the outer edge was determined to FWHM of up to 40.000 km at 8 $R_S$ distance. Such a large thickness is caused by inclined submicron sized particles, but this fact remained a major challenge for the dynamical ring models. Dikarev (1999) included plasma drag effects to the particle dynamics and he found a growth of the particles' semi-major axis. This allows the grains moving in less eccentric orbits to cover the full radial range of the ring which in turn increases the dust lifetime. Horányi et al. (2008) performed extensive numerical simulations of the long-term evolution of E ring particles taking into account sputtering and dust particle erosion, explaining a ring which could even extend to distances as far as the orbit of Titan.

In-situ investigations of the ring particles by the CDA detector helped to unveil some of the secrets of the E ring. Srama et al. (2006) summarized the first year of the in-situ investigations of Saturn's dust environment within the Titan orbit by the Cassini dust detector CDA. In two publications a comprehensive analysis of the CDA E ring data was given (Kempf et al., 2007; Postberg et al., 2007). Kempf et al. (2007, 2010) investigated the geometrical structure of the ring and analysed the ring particle size distribution, while

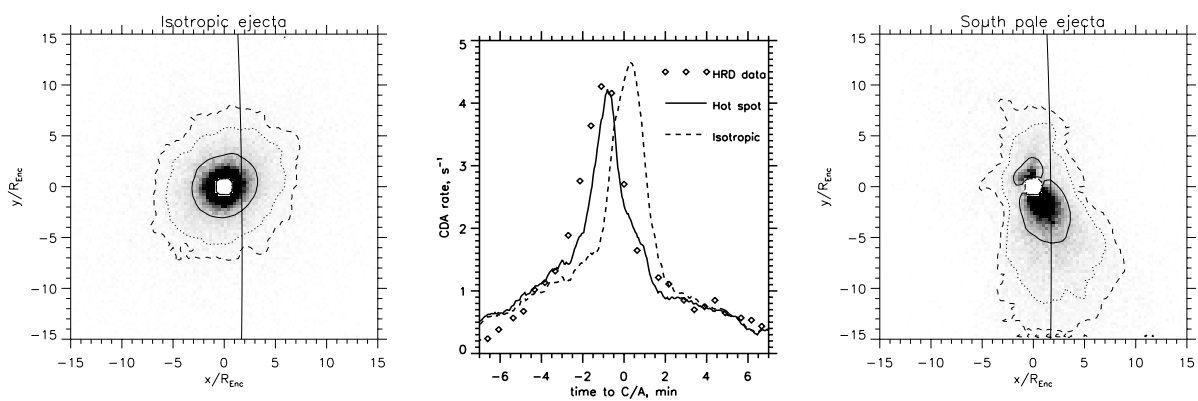

Figure 9: Discovery of an active dust source at the south pole of Enceladus as the dominant source of Saturn's E ring. Left: Simulated dust density of impact ejecta in the environment of the moon Enceladus. The line represents the trajectory of Cassini. Right: Simulated dust density of an active dust source at the south pole region of Enceladus. Middle: Measured dust rate by the CDA subsystem High Rate Detector (HRD) during the flyby and comparison with the two model predictions. The rate peaked almost one minute before the closest approach (C/A) to the Enceladus surface and is in agreement with an active source at the south pole (Spahn et al., 2006, Science).

Postberg et al. (2007) studied the composition of the ring particles.

**Enceladus ice geysers** Extensive research is focused to the dust production mechanism of the ring's main source, the icy moon Enceladus. Producing fresh dust particles by impacts of fast projectiles onto the moon's surface (the so-called impactor-ejecta process) has been accepted as the most effective process. In 1997 it was proposed that the E ring particles themselves constitute the main projectile source for replenishing the ring. However, energy considerations seem to favour the projectile flux being dominated by particles of interplanetary and interstellar origin. Angular distributions of ejecta generated by various populations of interplanetary dust particles (IDPs) were studied by Colwell (1993). Based on this Spahn et al. (1999) numerically studied the spatial ejecta distribution in the vicinity of Enceladus. Their predictions were used to optimise the CDA measurements during the close flybys of Enceladus in 2005.

This flyby on 14 July 2005 at an altitude of 170 km was supposed to become the most spectacular flyby of the Cassini mission. During this special observation CDA discovered a collimated jet of micron-sized dust particles emerging from Enceladus' south pole region (Fig. 9). Spahn et al. (2006b) inferred from comparison of numerical simulations with the CDA data that the particles emitted by the ice geysers at the south pole are the dominant source in the inner E ring. Recently, models for grain condensation and growth in the geyser channels were developed based on the Cassini data (Schmidt et al., 2008) which explain the observed particle speed and size distributions within the dust plume. Scientists assume a geyser venting area of approximately 200 m² with a total dust production rate of about 5 kg s$^{-1}$ of which ~10% escapes the satellites' gravity. On the other side, the mass loss of the E ring was calculated to be approximately 1 kg s$^{-1}$. Significant insight into the processes of the ice geysers below the icy crust of Enceladus were achieved by Postberg et al. (2009, 2011) by the chemical analysis of E ring grain compositions (see further below).

Another important source for ring particles are **secondary ejecta** produced at the icy surface of the embedded moons. Ring particles, interplanetary and interstellar dust are primary impactors producing ejecta with broad size and speed distributions. Ejecta particles with velocities higher than the escape speed of the individual moons become ring particles (Spahn et al., 2006, PSS, Krivov et al., 2003). For this process, they have to exceed speeds of 127 m s$^{-1}$ (Mimas), 205 m s$^{-1}$ (Enceladus), 338 m s$^{-1}$ (Tethys), 462 m s$^{-1}$ (Dione) and 592 m s$^{-1}$ (Rhea). E-ring impactors are a relevant sources for secondary ejecta for the inner moons and especially Enceladus and Tethys, where the ejecta mass production can be as high as 10 kg s$^{-1}$ or 3 kg s$^{-1}$, respectively. However, only 0.014 kg s$^{-1}$ are expected to escape from Rheas surface due to E ring impactors, whereas the surface ejecta of interplanetary meteoroids are leading to a mass flux of 0.019 kg s$^{-1}$ (Spahn et al., 2006, PSS). Therefore the outer moon Rhea with its encircled dust cloud is an excellent target in order to investigate the infalling interplanetary dust flux at Saturn. The low interplanetary dust flux at the distance of Saturn of

$F_{imp}$ = 1.8x10$^{-16}$ kg m$^{-2}$ s$^{-1}$ is „amplified" by the impactor ejecta process for icy surfaces according to M+ = $F_{imp}$Y S, with the cross section S of the source satellite, the mass production rate M+ and the yield Y = 2.64x10$^{-5}$ $m_{imp}^{0.23}$ $v_{imp}^{2.46}$ . $m_{imp}$ and $v_{imp}$ are the speed and mass of the impactors (IDPs), respectively, and Y is the fraction of the ejected mass to that of the projectile and was determined by laboratory experiments. Using a relative impact speed of the interplanetary grains with respect to the moon surface of 9500 m s$^{-1}$ and the dominating particle mass of 1x10$^{-8}$ kg, the yield Y becomes as high as 18.000 for Mimas and 7.500 for Rhea, making it possible to determine the interplanetary dust flux during one low altitude flyby of the moon Rhea (~100 km altitude, within the moon's Hill sphere). The Cassini flyby at Rhea on January 11th, 2011, occurred with an altitude of only 75 km and the CDA instrument pointing towards the dust RAM direction made it possible to acquire valuable data of highest quality for this purpose. The results of this ejecta cloud measurement are currently analysed and they will be published elsewhere.

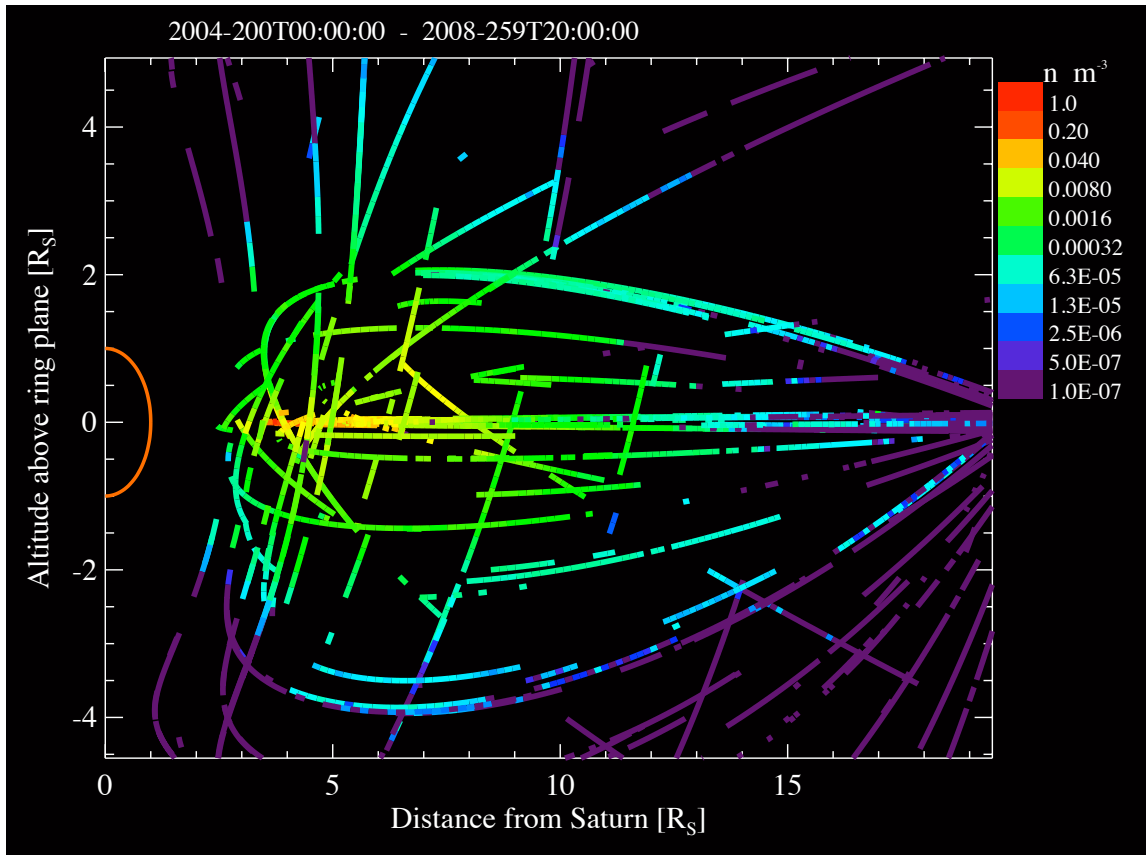

Figure 10: Global apparent dust density measured by CDA in the saturnian system. The densities are color coded along the Cassini trajectory in the time range of 2004-200 to 2008-259 (Cassini prime tour). High impact rates and densities are observed outside the optically measured E ring, which was defined between 3 to 9 RS. Enhanced dust densities are found as far as 250.000 km away from the ring plane and extends to radial distances of 20 RS and beyond.

Generally ejecta cloud measurements around Saturn's moons with Cassini-CDA are very difficult due to the high background density of the E ring particles. It was one of the major CDA goals to characterise the extension and sharpness of the edges of the largest planetary ring in our solar system. An in-situ instrument like CDA is approximately 1000 times more sensitive than optical remote sensing instruments to search and characterise faint dusty rings. However, in order to achieve a global view of the outer ring system, long integration times with adequate pointing are necessary. The fragmented observation timeline of Cassini (pointing changes), the small amount of pointing prime time for in-situ instruments and the spacecraft flight rules (forbidden Cassini attitudes primarily caused by IR instrument radiator heating) made this task a major challenge.

It was a big surprise to find dust particles at altitudes as high as 100.000 km above the ring plane and at Saturn distances of 7 $R_S$ early in the Cassini tour (e.g. 28 Oct 2004). The derived dust densities of particles larger than 0.5 μm reach values of 0.001 m$^{-3}$. The process to calculate dust densities from measured impact rates require the knowledge of the relative impact speeds, the projected sensitive instrument target area and of the instrument mass threshold, which itself is impact speed dependent. Furthermore, the relative impact speeds are calculated under the assumption of circular prograde orbits, but it is known, that E ring particles are also highly eccentric, leading to different relative impact speeds. Finally, the measured dust rates have to be corrected by further instrumental effects like the dead time, checkout times or threshold and triggers settings. Impact rates varied by seven orders of magnitude between 10.000 s$^{-1}$ close to Enceladus and 0.001 s$^{-1}$ in the outer ring regions. Therefore any densities given in this paragraph are apparent densities which might differ from real dust densities by a factor of 10. Taking the (conservative) CDA lower mass threshold $M_T$ of $M_T$ [kg] = 3.04x10$^{-13}$ v$^{-3.75}$ [km s$^{-1}$] into account (Srama, 2010), the calculated dust densities are giving the minimum densities of particles in a given volume. Grain size distributions are following potential laws such that tiny particles are more abundant than larger grains (Kempf, 2007). Assuming a typical relative impact speed of 8 km s$^{-1}$, the lower mass threshold would be 1.2x10$^{-16}$ kg. This mass corresponds to compact water ice particles with a diameter of 0.6 µm. Faster impact speeds decrease this mass threshold significantly: Impact speeds of 20 km s$^{-1}$ already produce enough impact charge for water ice grains with diameters above 200 nm to be detected.

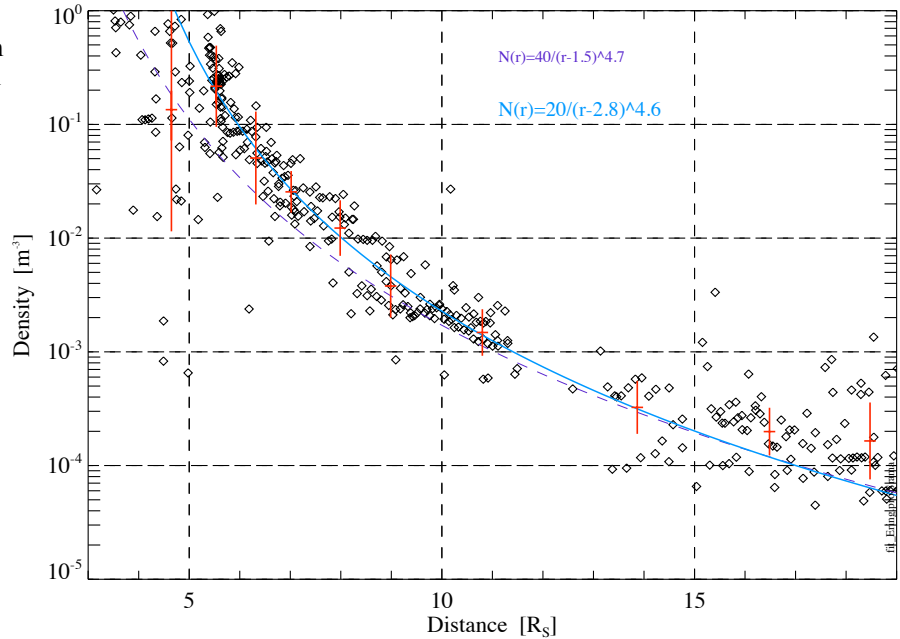

Figure 11: Radial density profile of Saturn's E ring in the ring plane derived from CDA data between 2004 and 2008 with a power law fit (blue line).

**E ring extension** How does the E ring of Saturn look like with the eyes of a sensitive in-situ instrument? For this purpose, the dust fluxes measured by CDA during the Cassini prime tour (DOY 200 in 2004 until DOY 260 in 2008) were processed and corrected for instrument settings, spacecraft pointing and relative impact speeds (assuming circular prograde particles). The result of this extensive data compilation is shown in Fig. 10. For this visualisation, the apparent dust density was plotted along the Cassini trajectory in a Saturn centered coordinate system in an edge-on view to the ring plane. The achieved pattern looks like pieces of a gift ribbon wrapping Saturn. Hidden or black areas along the track represent periods where CDA was not able to record meaningful data. This was primarily caused by long periods of inadequate pointing of Cassini (the dust RAM direction was not in the field-of-view of CDA). Furthermore, only counter data were taken into account in order to avoid any influence of the science data rate. Very tiny dust impacts like nanometre sized stream particles were not considered in this global overview, since this would totally change the picture: Stream particle data can cause CDA impact rates as high 1 s$^{-1}$ at distances of 30 $R_S$ or beyond and the derivation of a picture for the bound E ring particles would not be possible anymore. The highest dust densities in the inner Saturnian region (along

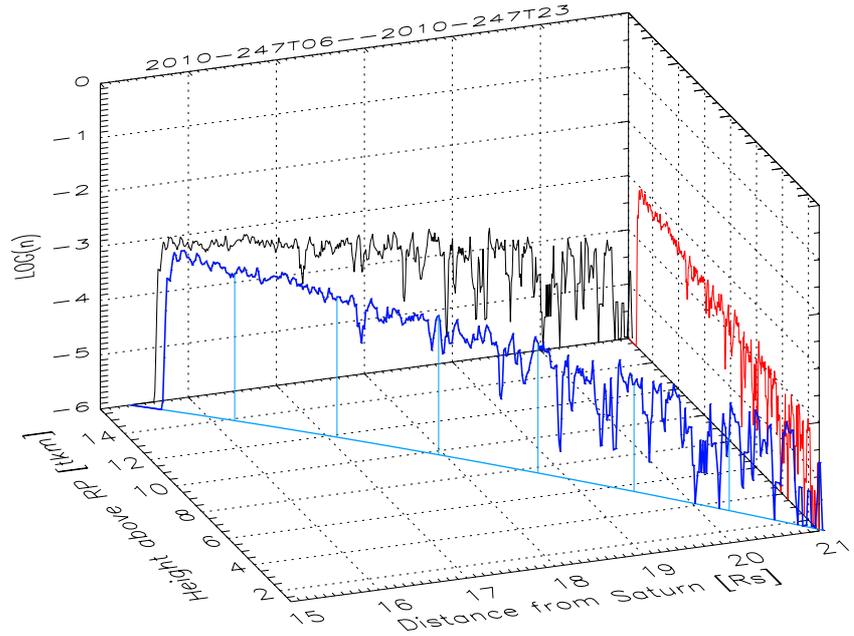

Figure 12 : Three-dimensional apparent dust density profile (blue) along Cassini's trajectory on DOY 247 in 2010 (orbit number 137). Two projections of the density profile are shown on the side panels (black and red curve). The vertical axis gives the dust density n in units of log(m−3). The density decreases from $1 \times 10^{-3}$ m$^{-3}$ at 15.5 R$_S$ down to $8 \times 10^{-5}$ m$^{-3}$ at 20 R$_S$ and an extension of the ring beyond Titan's orbit is visible. This discovery of an extended ring beyond Titan's orbit was confirmed in a later equatorial ring scan on DOY 290 in 2010 between 10 R$_S$ and 25 R$_S$ distance.

Enceladus orbit at 4 R$_S$) are a few particles per cubic meter and they are consistent with former modelling work. Surprising is certainly the high dust density inside Titan's orbit at 20 R$_S$ with altitudes up to 4 R$_S$ (250.000 km) from the ring plane. In the outer region, fluxes below $1 \times 10^{-6}$ m$^{-3}$ are measured.

The results shown in Fig. 10 were taken to derive the **radial profile** of the E ring close to the ring plane. The overall dust density is decreasing by three orders of magnitude between 6 R$_S$ and 16 R$_S$ radial distance (Fig. 11) and the density can be described by a power law of the type $n(r) = 20 (r - 2.8)^{-4.6}$. The applicable range of this law is constrained to the radial region between 6 and 15 R$_S$. At the inner boundary, the CDA dead time of 1 s limits the number of recorded particles (instrument saturation), and at the outer boundary the small number of impacts within one time bin leads to a higher noise. Starting at a distance of r =6 R$_S$ with a density of 0.095 m$^{-3}$ the density falls down to $2.0 \times 10^{-4}$ m$^{-3}$ at 15 R$_S$.

The **outer extension** of the ring is not well defined, but recent results indicate an outer boundary which lies even beyond Titan's orbit. Due to the fact that Cassini science is focused at Titan in the region around 20 R$_S$ distance, a special ring scan on DOY 247 in 2010 was integrated in the Cassini observation timeline. This observation with adequate dust RAM pointing covered a time period of approximately 16 hours between 15.5 R$_S$ and 21 R$_S$ and provided new insights into the outer structure of the E ring as shown in Fig. 12. Between 16 R$_S$ and 20 R$_S$ the dust density decreases by one order of magnitude starting at approximately $8 \times 10^{-4}$ m$^{-3}$ and ending at approximately $8 \times 10^{-5}$ m$^{-3}$. In this scenario, the mass threshold of CDA was $4.7 \times 10^{-16}$ kg (water ice grain with 1 µm diameter) for a relative impact speed of 5.6 km s$^{-1}$ for prograde circular dust grains at 20 R$_S$. Most particle are expected to move in eccentric orbits at this distance, leading to deviations in the relative impact speed and the lower mass threshold, respectively. The density along the trajectory seems to have some fine structure, but we have to take into account that at 20 R$_S$ only up to seven dust particles were counted within a time bin of 10 minutes leading to a high fluctuation of the calculated rate and density.

Nevertheless, this is the first in-situ measurement of Titan's dusty environment and we can take the calculated apparent density of 1 µm water ice particles to estimate the **influx of** **water** into Titan's atmosphere. Titan moves with 5.6 km s$^{-1}$ and the dust grains have a variety of eccentricities and we take a conservative relative dust impact speed of only 2 km s$^{-1}$. If we take now a Titan cross section of 2.1x10$^{13}$ m$^2$ we get a collection volume of 4.2x10$^{16}$ m$^3$ s$^{-1}$ meaning that 3.3x10$^{12}$ ice particles are raining down on Titan. This process would deliver a total amount of 1.6 g s$^{-1}$ of water into Titan's atmosphere. However, the assumed relative impact speed of 2 km s$^{-1}$ is probably too low, and the influx of tiny stream particles was also not considered in this estimation.

The overall amount of dust within Titan's orbit was very surprising. The questions that are immediately raised are: Do these particles belong to one big, faint and extended E ring, which was defined between 3 and 9 Rs earlier in the literature? Are there gaps or short scale density gradients in the extended dust environment? Are there dust density enhancements or different dust mass distributions along satellite orbits? The most likely answer is yes, there is one big E ring showing a very complex fine structure which is not explained until today.

## Summary

This paper is a first attempt to summarize selected scientific results of the Cosmic Dust Analyser onboard Cassini. CDA is the most advanced dust detector ever flown in interplanetary space. It combines subsystems to measure the dust particles properties like speed, mass, trajectory, composition and flux simultaneously and with high reliability. The scientific results reported in this paper are:

- The scientific success of CDA started already during the early cruise phase in 1999 by the discovery of the interstellar dust flux at one AU distance, which provided the basis for future planetary missions to measure interstellar dust in the vicinity of the Earth.
- After the Earth flyby in August 1999, CDA measured for the first time, electrostatic charges between 1 and 5 fC on interplanetary dust grains.
- For two interplanetary particles detected at 0.9 AU and 1.9 AU solar distance, the integrated time-of-flight mass spectrometer determined the elemental composition. These grains were predominantly iron with surprisingly few traces of silicates.
- CDA provided even the chemical composition of Jovian dust stream particles: nanometre sized particles originating from Jupiter's moon Io escaping from the Jovian system with speeds up to 400 km s$^{-1}$. These grains primarily consist out of NaCl with possible contributions of sulfur and silicates (Postberg et al., 2006).
- CDA discovered dust streams escaping from the Saturnian system. The first stream was detected 70 Mkm away from Saturn. The grains are composed out of silicate material and about 30% of them have mantles consisting of water ice and/or clathrate hydrates of ammonia. Source regions for these grains, which are basically slower and smaller than their Jovian counterparts, are the inner A ring and the E ring outside of Dione's orbit. The streams are correlated with properties of the IMF. Furthermore, CIRs play a major role in the stream flux and directionality.
- The HRD subsystem of CDA detected the outer rim of Saturn's G ring in 2005. Unfortunately, this event probably damaged one of the HRD foil sensors leading to a higher noise rate. Fortunately, this detector remained functional during the ongoing tour.
- CDA was one of the major instruments discovering the dust plumes at the south pole of Enceladus during the flyby in July 2005. The active ice volcanoes eject submicron and micron-sized particles with a mass flux of approximately 5 kg s$^{-1}$ of which 10% can escape replenishing the faint E ring. The dust plumes of Enceladus became the dominant dust source of the E ring. Ejecta particles generated at the surfaces of the outer moons by interplanetary or E ring impactors are a second source for E ring particles.
- During the global characterisation of the E

ring CDA discovered an unexpected wide vertical and radial extension. The vertical extension is by a factor of 10 higher than observed by optical instruments. The radial extension reaches the orbit of Titan and beyond. This lead to a new understanding of the dynamics of the outer ring.

The major discoveries of CDA which are not part of this paper are:
- Measurements of the ice grain potential in Saturn's E ring which lead to a new plasma model of the magnetosphere (Kempf et al., 2006, PSS)
- The in-situ characterisation of the E ring dust particles in the vicinity of Enceladus was published by Kempf et al. (2008, Icarus). For the first time, in-situ profiles of the vertical and radial ring profile as well as particle mass distributions were given. Based on these results, a dust injection model of the ice volcanoes at the surface of Enceladus was developed (Schmidt et al., 2008, Kempf et al., 2010). This dust injection model was the basis for a new dynamic model explaining how the Enceladus dust plume feeds Saturn's E ring (Kempf et al., 2010). The dust plume model predicted patterns of snow layers at the surface of Enceladus which are in agreement with recent observations.
- Planetologists concluded from CDA data the existence of an ocean of liquid water below the icy crust of Enceladus. This scientific highlight is based on the observation of sodium salts in the E-ring ice grains and was reported by Postberg et al. (2009, 2011).

The long list of discoveries still leaves room for future exploration. Today, CDA science observations focus on an in-depth analysis of ice grain compositions close to Enceladus, seasonal effects of Saturn's ring system, the determination of the interplanetary and interstellar dust flux at Saturn, the search for retrograde dust populations and possible asymmetries in the ring system, a global understanding of the grain composition in Saturn's E ring and the dust populations outside of Titan's orbit.

## Acknowledgements

The Cassini Cosmic Dust Analyser science planning, operations and science analysis is supported by the Deutsches Zentrum für Luft- und Raumfahrt e. V. (DLR) on behalf of the BMWi under the project grant 50 OH 1103.